\documentclass[
 article ]
  {aipproc}

\layoutstyle{6x9}

\usepackage{graphicx,epsf,amssymb}

\newcommand{\be}{\begin{equation}} 
\newcommand{\ee}{\end{equation}}

\newcommand{\bea}{\begin{eqnarray}} 
\newcommand{\eea}{\end{eqnarray}} 
\newcommand{\Tr}{{\rm Tr}}

\def\V{V}

\newif\ifdraft
\drafttrue
\newif\ifpreprint
\preprinttrue

\def\fig#1{fig.~{\ref{#1}}}

\def\pol{\varepsilon}
\def\Tr{\, {\rm Tr}}

\def\spa#1.#2{\left\langle#1\,#2\right\rangle}
\def\spb#1.#2{\left[#1\,#2\right]}
\def\sand#1.#2.#3{%
\left\langle\smash{#1}{\vphantom1}^{-}\right|{#2}%
\left|\smash{#3}{\vphantom1}^{-}\right\rangle}
\def\sandp#1.#2.#3{%
\left\langle\smash{#1}{\vphantom1}^{-}\right|{#2}%
\left|\smash{#3}{\vphantom1}^{+}\right\rangle}
\def\sandpp#1.#2.#3{%
\left\langle\smash{#1}{\vphantom1}^{+}\right|{#2}%
\left|\smash{#3}{\vphantom1}^{+}\right\rangle}
\def\sandpm#1.#2.#3{%
\left\langle\smash{#1}{\vphantom1}^{+}\right|{#2}%
\left|\smash{#3}{\vphantom1}^{-}\right\rangle}
\def\sandmp#1.#2.#3{%
\left\langle\smash{#1}{\vphantom1}^{-}\right|{#2}%
\left|\smash{#3}{\vphantom1}^{+}\right\rangle}
\def\sandmm#1.#2.#3{%
\left\langle\smash{#1}{\vphantom1}^{-}\right|{#2}%
\left|\smash{#3}{\vphantom1}^{-}\right\rangle}
\def\spab#1.#2.#3{\sandmm#1.#2.#3}
\def\spba#1.#2.#3{\sandpp#1.#2.#3}
\def\spaa#1.#2.#3.#4{\sandmp#1.{#2#3}.#4}
\def\spbb#1.#2.#3.#4{\sandpm#1.{#2#3}.#4}
\def\spash#1.#2{\spa{\smash{#1}}.{\smash{#2}}}

\newbox\charbox
\newbox\slabox
\def\s#1{{      % Feynman slash
        \setbox\charbox=\hbox{$#1$}
        \setbox\slabox=\hbox{$/$}
        \dimen\charbox=\ht\slabox
        \advance\dimen\charbox by -\dp\slabox
        \advance\dimen\charbox by -\ht\charbox
        \advance\dimen\charbox by \dp\charbox
        \divide\dimen\charbox by 2
        \raise-\dimen\charbox\hbox to \wd\charbox{\hss/\hss}
        \llap{$#1$}
}}

\def\eqn#1{eq.~(\ref{#1})}

\def\qb{{\overline {\kern-0.7pt q\kern -0.7pt}}}

\def\ep{{e^+}}
\def\em{{e^-}}

\def\ib{{\bar\imath}}

\def\sign{{\mathop{\rm sign}\nolimits}}

\def\sandp#1.#2.#3{%
\left\langle\smash{#1}{\vphantom1}^{+}\right|{#2}%
\left|\smash{#3}{\vphantom1}^{+}\right\rangle}
\def\ksl{\s{k}}

\def\proj{{\flat}}

\newbox\ourfigbox
\def\SizedFigureWithCaption#1#2#3{%
\setbox\ourfigbox
  \hbox{\hss\epsfxsize #1 \epsfbox{#2}\hss}
\hbox to \wd\ourfigbox{\vbox{\noindent\copy\ourfigbox\break
\vskip -6mm      \hbox to \wd\ourfigbox{\hss#3\hss}}}
}

\def\spa#1.#2{\left\langle#1\,#2\right\rangle}
\def\spb#1.#2{\left[#1\,#2\right]}
\def\lor#1.#2{\left(#1\,#2\right)}
\def\sand#1.#2.#3{%
\left\langle\smash{#1}{\vphantom1}^{-}\right|{#2}%
\left|\smash{#3}{\vphantom1}^{-}\right\rangle}
\def\sandpp#1.#2.#3{%
\left\langle\smash{#1}{\vphantom1}^{+}\right|{#2}%
\left|\smash{#3}{\vphantom1}^{+}\right\rangle}
\def\sandpm#1.#2.#3{%
\left\langle\smash{#1}{\vphantom1}^{+}\right|{#2}%
\left|\smash{#3}{\vphantom1}^{-}\right\rangle}
\def\sandmp#1.#2.#3{%
\left\langle\smash{#1}{\vphantom1}^{-}\right|{#2}%
\left|\smash{#3}{\vphantom1}^{+}\right\rangle}

\begin{document}

\hfill{UCLA/05/TEP/28}

\title{CSW Diagrams and Electroweak Vector Bosons
{\footnote{Presented at the workshop {\it QCD at Work, 2005},
Conversano (BA), Italy, June 16-20, 2005. Research supported by the US Department of
Energy under contract DE--FG03--91ER40662.}}}



\author{Pierpaolo Mastrolia}{
  address={Department of Physics and Astronomy, UCLA
           Los Angeles, CA 90095--1547, USA}}

\begin{abstract}
Based on the joined work performed together with Z. Bern, D. Forde, and D. Kosower 
\cite{Bern:2004ba}, in this talk it is recalled the (twistor-motivated) diagrammatic formalism
describing tree-level scattering amplitudes presented by Cachazo, Svr\v{c}ek and Witten, 
and it is discussed an extension of the vertices and accompaining rules 
to the construction of vector-boson currents coupling to an arbitrary
source. 
\end{abstract}

\maketitle

%%%%%%%%%%%%%%%%%%%%%%%%%%%%%%%%%%%%%%%%%%%%
%% MAINMATTER
%%%%%%%%%%%%%%%%%%%%%%%%%%%%%%%%%%%%%%%%%%%%

\section{Introduction}
\label{IntroSection}

The computation of amplitudes for QCD and mixed electroweak-QCD 
processes is an important part of a physics program at modern-day
colliders, given their important role
as experimentally distinctive probes of new physics.

In less than a couple of years, 
the progress in the evaluation of scattering processes has received a strong
boost, due to a better understending of the analytic structure of scattering
amplitudes. Stimulated by Witten's realization \cite{WittenTopologicalString}
that tree-level gluon amplitudes, once transformed in twistor space \cite{Penrose},
have a simple geometrical description,
Cachazo, Svr\v{c}ek and Witten (CSW)~\cite{CSW} proposed a powerful set of new computational
rules to deal with many particles scattering amplitudes in QCD. 
Only recently, Risager \cite{CSWdemo} has found that the CSW approach can be
understood as a particular class of a more general set of {\it on-shell
  recurrence relations} for amplitudes, which meanwhile had been introduced by Britto,
Cachazo, Feng and Witten (BCFW) \cite{BCFW}.

The CSW rules are of interest in their own right for
tree-level computations. Their efficiency for gluonic amplitudes, 
was soon extended to account for massless external fermions \cite{Khoze}
and Higgs boson coupled to
QCD via a massive top-quark loop (in the infinite-mass limit) \cite{LanceHiggs},
and improved when recast in a recursive form~\cite{MHVRecursive}.
They also allow great simplification in loop calculations~\cite{TwistorLoop}.  

In this talk is described the extension of the CSW construction to include building
blocks for mixed QCD-electroweak amplitudes, by providing a 
construction of vector currents, 
which may in principle be coupled to an {\it arbitrary\/} source.
We will focus on the case of coupling a process involving one quark
pair and any number of gluons to one colorless off-shell vector boson.
The key idea in the
construction is to introduce a new set of basic vertices coupling to
the off-shell vector boson, having either one or no negative-helicity
gluons. The rules for combining them into new currents with additional
negative-helicity legs are then in fact the same as those of CSW.

%%%%%%%%%%%%%%%%%%%%%%%%%%%%

\section{Color Decompositions}
\label{ReviewSection}

Color decompositions~\cite{TreeColor,TreeReview} allows the disentangling of
the the gauge group factors 
and the pure kinematical terms, namely {\it partial amplitudes}, in the
full momentum-space amplitudes. 
For example, the tree-level $n$-gluon amplitude ${\cal A}_n$ has the 
color decomposition,
\begin{equation}
{\cal A}_n(1,2, \ldots, n ) = 
\sum_{\sigma \in S_n/Z_n} \Tr(T^{a_{\sigma(1)}}\cdots T^{a_{\sigma(n)}})\,
A_n(\sigma(1),\ldots,
      \sigma(n)))\,,
\label{TreeColorDecomposition}
\end{equation}
where $S_n/Z_n$ is the group of non-cyclic permutations
on $n$ symbols, and $j$ denotes the $j$-th gluon and
its associated momentum. 
We use the color normalization $\Tr(T^a T^b) = \delta^{ab}$. 
Similar decompositions hold for cases involving quarks. In general, it is 
more convenient to calculate the partial amplitudes than the
entire amplitude at once.

The cases in which we are interested here involve colorless vector
bosons.  Single massive
vector boson exchange is easily obtained from pure QCD amplitudes
(which are directly calculable from CSW diagrams).  For example, for
$\ep {\em} \rightarrow \gamma^* \rightarrow q \qb + n$ gluons, where
$\gamma^*$ represents an off-shell photon, the amplitude reduces to
\begin{eqnarray}
%\null\hskip -.5 cm 
 {\cal A}_n( 1_\ep, 2_{\em}, 3_q, 4, 5, \ldots, (n-1),  n_\qb)  & = & 
-2 e^2 Q^q g^{n-2} 
 \times \sum_{\sigma \in S_{n-4}} (T^{a_\sigma(4)} \cdots
    T^{a_\sigma(n-1)})_{i_3}^{\; \ib_n} \nonumber \\
&& \hskip -.5 cm \null
 \times \ A_n( 1_\ep, 2_{\em}, 3_q, \sigma(4), \ldots, 
    \sigma(n-1),  n_\qb) \ , \qquad 
\label{eeZPartons}
\end{eqnarray}
where we use an all outgoing momentum convention.
The particle labels $q, \qb, {\em}, \ep$ stand for quarks, anti-quarks,
electrons and positrons, while legs without labels, for gluons.
The off-shell photon, is internal to the amplitude and exchanged
between the lepton pair and the quark pair.

To convert the exchanged photon to an electroweak vector
boson, for describing 
$ e^+ e^- \rightarrow Z,\gamma^* \rightarrow \qb q + n g$,
one may adjust the coupling and modify the photon
kinematic pole to account for an unstable massive
particle ~\cite{BGKVectorBoson,DKS}. More generally, one may convert gluons to
photons, purely by group 
theoretic rearrangements \cite{QQGGG}.  However, in general, it is not
possible to then convert the photonic amplitudes to ones involving
electroweak vector bosons since vector bosons have non-abelian self
interactions which photons do not.  

A purpose of this talk is to see how constructing an appropriate off-shell
continuation so that the CSW {\it diagrammatricks}, originary introduced to
describe pure gluon scattering, can be applied to such
cases as well.

%%%%%%%%%%%%%%%%%%%%%%%%%%%%%%%%%%%%%%%%%%%%%

\section{CSW Diagrams}
\label{CSWSection}

\def\vo{\vphantom{1}}
The CSW construction~\cite{CSW} builds amplitudes out of vertices
which are off-shell continuations of the Parke--Taylor amplitudes
\cite{ParkeTaylor, Recurrence}.
These amplitudes, with two negative-helicity
gluons and any number of positive-helicity ones, are the maximally
helicity-violating (MHV) non-vanishing tree-level amplitudes in a
gauge theory.
In the spinor helicity~\cite{SpinorHelicity,XZC,TreeReview}
notation, they are,
\begin{equation}
A_n(1^+,\ldots,m_1^{-},\ldots,m_2^{-},\ldots,n^+) = 
  i {\spa{m_1}.{m_2}^4\over \spa1.2\spa2.3\cdots \spa{(n\!-\!1)}.n\spa{n}.1},
\end{equation}
where the two negative-helicity gluons are labeled
$m_{1,2}$. In this equation,
$\spa{i}.{j} = \spa{k_i}.{k_j}$.
We follow the standard spinor normalizations
$\spb{i}.{j} = \sign(k_i^0 k_j^0)\spa{j}.{i}^*$ and
$\spa{i}.{j}\spb{j}.{i} = 2 k_i\cdot k_j$.
With our conventions all particle momenta are taken to be outgoing.

The remaining MHV fermionic amplitudes needed for our discussion
of vector boson currents are,
\begin{eqnarray}
&& A(1_q^+,2^+,3^+, \ldots, i^-, \ldots, (n-2)^+, (n-1)_\qb^-, n^+) =
i {\spa{i}.{1} \spa{i,\,}.{n-1}^3 \over
              \spa1.2 \spa2.3 \spa3.4 \cdots \spa{n}.1}\,, 
\label{MHVtwoquarkphotonD}\\
&& A(1_q^+,2^+,3^+, \ldots, (n-2)^+, (n-1)_\qb^-, n^-) =
i {\spa{n}.{1} \spa{n-1,\,}.n^3 \over
              \spa1.2 \spa2.3 \spa3.4 \cdots \spa{n}.1}\,, 
\label{MHVtwoquarkphotonB} \\
&& A(1_{\qb'}^-,2_{q'}^+, 
    3_q^+, 4^+, \ldots, (n-1)^+, n_\qb^-) 
=  -i {\spa{1}.n^2 \over \spa1.2 \spa3.4 \spa4.5
           \cdots \spa{(n-1)}.{n}} \,. \hskip 1 cm 
\label{MHVfourquarkB}
\end{eqnarray}
The last equation gives the color-ordered amplitude appearing in
\eqn{eeZPartons} after relabeling $q' \rightarrow {\em} $ and
$\qb'\rightarrow \ep$.  

In the CSW construction a particular off-shell continuation of these
amplitudes, $A$, is an MHV vertex, $V$.  The original CSW prescription for the
off-shell continuation of a momentum $k_j$ amounts to replacing
$
%\begin{equation}
\spa{j}.{j'}  \longrightarrow \spb{\eta}.j \spa{j}.{j'}
 \longrightarrow \sandp{\eta}.{\ksl_j}.{j'},
%\label{CSWOffShell}
%\end{equation}
$
where $\eta$ is an arbitrary light-like reference vector, in the
Parke-Taylor formula.  The extra factors introduced in this off-shell
continuation cancel when sewing together vertices to obtain an
on-shell amplitude.  As shown by CSW~\cite{CSW}, on-shell amplitudes
are in fact independent of the choice of $\eta$, implying that the
sum over MHV diagrams is Lorentz invariant.

In our construction we use an alternative, but equivalent way of going
off-shell~\cite{NMHVTree,MHVRecursive}.  We instead decompose an
off-shell momentum $K$ into a sum of two massless momenta, where one
is proportional to the auxiliary light-cone reference momentum $\eta$ (with
$\eta^2 = 0$),
\begin{equation}
K = K^\proj + \zeta(K) \eta \, .
\end{equation}
The constraint $(K^\proj)^2 = 0$ yields
%\begin{equation}
$
\zeta(K) =  K^2/(2 \eta\cdot K).
$
%\end{equation}
%
If $K$ goes on shell, $\zeta$ vanishes.  Also, if two off-shell vectors
sum to zero, $K_1+K_2=0$, then so do the corresponding $k^\proj$s.
The prescription for continuing MHV amplitudes or vertices off shell is 
to replace,
\begin{equation}
\spa{j}.{j'}\rightarrow \spa{\smash{j^\proj}}.{j'},
\label{OffShellPrescription}
\end{equation}
when $k_j$ is taken off shell.  In the on-shell limit, $\zeta(K)$ vanishes
and $k_j^\proj \rightarrow k_j$.  Although equivalent to the original CSW
prescription, it is a bit more convenient to implement.  In particular,
there are no extra factors associated with going off-shell 
and the MHV vertices carry the same dimensions as amplitudes.

The CSW construction replaces ordinary Feynman diagrams with diagrams
built out of MHV vertices and ordinary propagators.  Each vertex has
exactly two lines carrying negative helicity (which may be on or off
shell), and at least one line carrying positive helicity.  The
propagator takes the simple form $i/K^2$, because the physical state
projector is effectively supplied by the vertices.  
For example, with this notation an all-gluon vertex would
be,
\begin{eqnarray}
&& V(1^+,\ldots,m_1^{-},(m_1\!+\!1)\vo^{+},\ldots, n,
          K\vo^-)\ = i\, 
{\spash{m_1}.{K^\proj}^4 \over \spa1.2\spa2.3\cdots 
      \spash{n}.{K^\proj} 
      \spash{K^\proj}.1} \,.
\label{Vertices}
\end{eqnarray}
%

%% The fermionic vertex reads \footnote{For the issue of the sign of 
%%   the fermionic vertex,  we remind the reader to the
%%   paper \cite{Bern:2004ba}.},
%% \begin{eqnarray}
%% && \hskip -1.0 cm 
%% V(1^+,\ldots,f_1^+,\ldots,m_1^{-},(m_1\!+\!1)\vo^{+},\ldots, n,
%%           f_2\vo^-) 
%% =
%%  i \, \sign(k_{f_2}^0)\, 
%% {\spa{m_1}. {f_2}^3 \spa{m_1}.{f_1}
%%   \over \spa1.2\spa2.3\cdots 
%%      \spa{n}.{f_2} \spa{f_2}.1} \,, \nonumber \\
%% \label{FermionicVertex}
%% \end{eqnarray}
%
%% where as usual legs are continued off shell by replacing $k_j$ with
%% $k_j^\proj$, and where $-f_2$ denotes $-k_{f_2}$. 

The CSW rules then instruct us to write down all tree diagrams with
MHV vertices, subject to the constraints that each vertex has exactly
two negative-helicity gluons and at least one positive-helicity gluon
attached, and that each propagator connects legs of opposite helicity.
For amplitudes with two negative-helicity gluons, the vertex with all
legs taken on shell is then the amplitude.  For each additional
negative-helicity gluon, we must add a vertex and a propagator.  The
number of vertices is thus the number of negative-helicity gluons,
less one.

%%%%%%%%%%%%%%%%%%%%
%FIGURE
%
\begin{figure}[t]
\centerline{\epsfxsize 3.5 truein \epsfbox{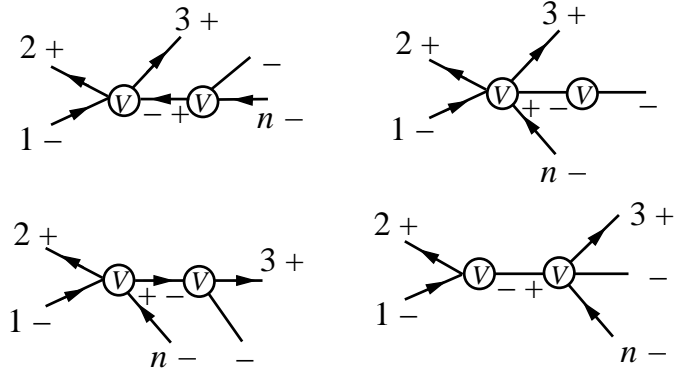}}
\caption{The stripped diagrams for 3 minus helicity amplitudes with
vector boson exchange between two fermion pairs.  Legs 1 and 2
correspond to the leptons and the legs 3 and $n$ to the quarks.
Lines with arrows represent quarks and those without arrows represent
either vector bosons or gluons.}
\label{A4fmmmFigure}
\end{figure}
%%%%%%%%%%%%%%%%%%%%%%

As a simple example we may use the CSW rules to construct
next-to-MHV (NMHV) partial amplitudes needed for the process $e^+ e^-
\rightarrow \gamma^*, Z, W \rightarrow q \bar q +ng$.
The `stripped diagrams' (where all the positive helicity gluons are not indicated) 
for this process
are shown in \fig{A4fmmmFigure}.  
Dressing the diagrams with the positive helicity gluon
legs between $q$ and $\qb$ in the color ordering leads to

{\footnotesize
\begin{eqnarray}
&& \hskip -.4 cm 
A(1_{q'}^-, 2_{\qb'}^+,3_q^+, 4^+,5^+, \ldots, (n-1)^-, n_\qb^-)\nonumber\\
&& \hskip 0.5cm 
= \sum_{j = 4}^{n-1}
  V(1_{q'}^-, 2_{\qb'}^+,3_q^+,4^+,5^+, \ldots, (j-1)^+, 
             (-K_{1\ldots (j-1)})_\qb^{-}) \, {i \over K_{1\ldots (j-1)}^2} 
             \nonumber \\
&& \hskip 2.7 cm \null \times
         V((-K_{j\ldots n})_q^{+}, j^+, \ldots, (n-2)^+, 
               (n-1)^-, n_\qb^-) \nonumber \\
&&  \hskip 0.6 cm \null
+  \sum_{j=4}^{n-2}  V(1_{q'}^-, 2_{\qb'}^+,3_q^+,4^+,5^+, \ldots, (j-1)^+, 
                   (-K_{n 1\ldots (j-1)})^{+}, n_\qb^-) \, 
                             {i \over K_{n 1\ldots (j-1)}^2} \nonumber\\
&& \hskip 2.7 cm \null \times
         V((-K_{j \ldots (n-1)})^{-}, j^+, \ldots, (n-2)^+, 
               (n-1)^-)  \nonumber \\
&&  \hskip 0.6 cm \null
+  V(1_{q'}^-, 2_{\qb'}^+, (-K_{n, 1, 2})_q^{+}, n_\qb^-) \, 
                                {i \over K_{n12}^2} \, 
%%  \\
%% && \hskip 2.7 cm \null \times
     V(3_q^+, 4^+, \ldots, (n-2)^+,  (n-1)^-,  
                 (-K_{3 \ldots (n-1)})_\qb^{-}) 
                \nonumber \\
&&  \hskip 0.6 cm \null
 +  V(1_{q'}^-, 2_{\qb'}^+, (-K_{1 2})^{-} ) \, 
                                {i \over K_{12}^2}  
%\nonumber\\
%&& \hskip 2.7 cm \null \times  
  V(3_q^+, 4^+, \ldots, (n-2)^+,  (n-1)^-,  n_\qb^-, 
         (-K_{3 \ldots n})^{+} ) \, ,
\label{NMHVVectorBosonExchange}
\end{eqnarray}
}%% end-footnotesize
where $K_{i\ldots j} = k_i + k_{i+1} + \cdots + k_j$.  
Renaming $\qb',q' \rightarrow \ep, \em$ 
gives the partial amplitudes appearing in the
vector boson exchange amplitudes (\ref{eeZPartons}).  

%%%%%%%%%%%%%%%%%%%%%%

\section{MHV Vertices for Vector Boson Currents}
\label{CurrentsSection}

In this section we generalize the CSW construction to allow couplings
to arbitrary sources.  We focus on the phenomenologically interesting
case of vector boson currents, though our construction of currents is
applicable more generally.  

An important application of these currents is that they allow us to
couple the electroweak theory to QCD, while taking full advantage of the
CSW formalism on the QCD side.  The currents satisfy a similar color
decomposition as the photon exchange amplitude~(\ref{eeZPartons}),
\begin{eqnarray}
\null\hskip -.5 cm 
 {\cal J}_\mu( 1_q, 2, 3, \ldots, (n-1),  n_\qb; P_\V)  & = & 
g_\V\, g^{n}\sum_{\sigma \in S_{n-2}} (T^{a_\sigma(2)} T^{a_\sigma(3)} \cdots
    T^{a_\sigma(n-1)})_{i_1}^{\; \ib_n} \nonumber \\
&& \hskip .1 cm \null
 \times  J_\mu( 1_q, \sigma(2), \sigma(3), \ldots, 
    \sigma(n-1),  n_\qb ; P_\V) \,,
\label{CurrentColor}
\end{eqnarray}
where $g_\V$ is the appropriate coupling for a vector boson $\V=\gamma^*,Z,W$
and $P_\V$ is the momentum carried by the vector boson.
Hence we need consider only the partial currents $J_\mu$ in much 
the same way that we need only consider color-ordered partial amplitudes.

We start by defining two currents that will serve
as new basic vertices for obtaining general vector boson currents:
\begin{enumerate}

\item A vector-boson current with $n$ gluon emissions, all of positive
helicity
\begin{eqnarray}
\null \hskip -.5 cm 
J^\mu(1_q^-,2^+, \ldots, (n-1)^+, n_\qb^+; P_\V) 
  &=& -{i \over \sqrt{2}} {\sandmp{(-1)}.{\gamma^\mu \s P_\V}.{(-1)}
 \over \spa{(-1)}.2 \spa2.3 \ldots \spa{(n-1)}.{n} } \nonumber\\
 &=&  c_+ \, \pol^{(+)\mu} (P_\V^\proj,\eta)
     +c_- \, \pol^{(-)\mu} (P_\V^\proj,\eta) \nonumber \\
&& \hskip .5 cm \null 
  + c_L \, \biggl( P_\V^\mu - {P_\V^2\over \eta\cdot P_\V} \eta^\mu\biggl)\,, 
\hskip 3 cm 
\label{BasicFermionicCurrent}
\end{eqnarray}
where $P_\V = -K_{1\ldots n}$ by momentum conservation, where `$-1$'
as a spinor argument denotes $-k_1$, and where
\begin{eqnarray}
c_+ &=& -V^{\rm MHV}(1_\qb^-,\ldots,n_q^+;P_\V^-)\,,\nonumber\\
c_- &=& V^{\rm MHV}(1_\qb^-,\ldots,n_q^+;P_\V^-) \,
             {\spa1.{\eta}^2 P_\V^2\over
              \spash{\eta}.{P_\V^\proj}^2\spash1.{P_\V^\proj}^2}\,,
\label{BasicFermionCoefficients}\\
c_L &=& V^{\rm MHV}(1_\qb^-,\ldots,n_q^+;P_\V^-) \,
{\sqrt2\spa1.{\eta}
 \over\spash{\eta}.{P_\V^\proj}\spash1.{P_\V^\proj}} \,.\nonumber
\end{eqnarray}
The vertex $V^{\rm MHV}$ is simply a CSW vertex for one photon, one
quark pair, and $n-2$ gluons, obtained by fermionic phase adjustments
from the amplitude in \eqn{MHVtwoquarkphotonB}.
As with the basic CSW vertices, when any colored leg $j$ is taken off shell, the $k_j$
argument to all spinor products or spinor strings must be replaced by
$k_j^\proj$.  

\item A purely bosonic basic current emitting a single vector
state,
\begin{equation}
J^\mu((-P_\V)^-;P_\V) 
  = {i \over \sqrt{2}} {\sandpp{\eta}.{\gamma^\mu}.{P_\V^\proj}
                         \over \spb{P_\V^\proj}.{\eta}} P_\V^2
   = i \, \pol^{(-)\mu}(P_\V^\proj, \eta) P_\V^2 \,.
\label{BasicGluonicCurrent}
\end{equation}
\end{enumerate}
The first of these is the vector-boson current for positive
helicity gluons~\cite{BGKVectorBoson}. The second is just a negative
helicity polarization vector with reference momentum taken to be the
CSW reference momentum.

The polarizations in the above equations are defined using
the spinor helicity method and are given by~\cite{XZC} 
\begin{equation}
\pol_\mu^{(+)}(k, r) = 
 {1\over \sqrt{2}} {\sandmm{r}.{\gamma^\mu}.{k} \over \spa{r}.{k}}\,,
\hskip 2 cm 
\pol_\mu^{(-)}(k, r) = 
 {1\over \sqrt{2}} {\sandpp{r}.{\gamma^\mu}.{k} \over \spb{k}.{r}}\,,
\end{equation}
where $r$ is a null reference momentum.

We take the currents (\ref{BasicFermionicCurrent}) and
(\ref{BasicGluonicCurrent}) to act as vertices, using the same CSW
prescriptions (\ref{OffShellPrescription}) as
used for defining vertices from MHV amplitudes.

%%%%%%%%% FIGURE %%%%%%%%%%%%%%%
\begin{figure}[t]
%\centerline{\epsfxsize 6 truein \epsfbox{Jfmm.eps}}
  \includegraphics[height=.09\textheight]{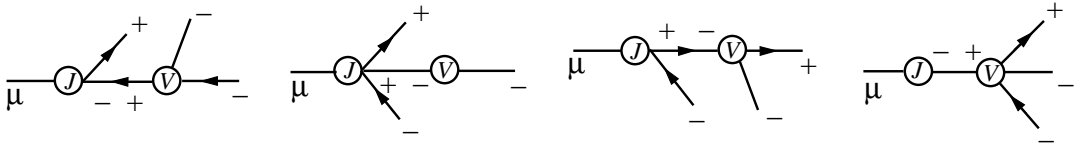}
\caption{
The NMHV vector boson current in terms of diagrams where
positive helicity gluon lines have been stripped.
}
\label{JfmmFigure}
\end{figure}
%%%%%%%%%%%%%%%%%%%%%%%%%%%%%%%%

To illustrate the construction of a current with more negative
helicities, consider the NMHV vector boson current,
$J_{\mu}(1_q^+,2^+,\ldots,(n-2)^+,(n-1)^-,n^{-}_{\qb};P_\V)$
where the negative helicity legs are nearest neighbors in the
color ordering.  The CSW diagrams for this current may be organized
using the four diagrams shown in \fig{JfmmFigure},
where the positive helicity gluon legs have all been stripped away.
Inserting back the positive helicity gluon legs, leads to the following
expression for this NMHV vector boson current,

{\footnotesize
\begin{eqnarray}
&&
\hskip -0.4 cm 
J_{\mu}(1_q^+,2^+,\ldots,(n-2)^+,(n-1)^-,n^{-}_{\qb}; P_\V)
\nonumber\\
&& \hskip 1 cm 
=\sum_{j=2}^{n-1}J_{\mu}(1_q^+,2^+,\ldots,(j-1)^+,
  (K_{j\ldots n})^{-}_{\qb}; P_\V)
  \frac{i}{K_{j \ldots n}^2}
\nonumber \\
&& \hskip 2 cm \null \times
V( (-K_{j\ldots n})_{q}^{+}, 
    j^+,\ldots,(n-2)^+,(n-1)^-,n^-_{\qb})
\nonumber \\
&&\hskip 1.2 cm \null 
+\sum_{j=2}^{n-2}J_{\mu}(1_q^+,2^+,\ldots,(j-1)^+,
               (K_{j\ldots (n-1)})^{+},n^-_{\qb}; P_\V )
\frac{i}{K_{j\ldots (n-1)}^2}
\nonumber \\
&& \hskip 2.5 cm \times \null 
V((-K_{j \ldots (n-1)})^{ -} ,j^+,\ldots,(n-2)^+,(n-1)^-)
\nonumber \\
&& \hskip 1.2 cm \null
+J_{\mu}((K_{1 \ldots (n-1)})^{ +}_{q},n^-_{\qb}; P_\V)
\frac{i}{K_{1 \ldots (n-1)}^2}
%% \\
%% &&  \hskip 2.5 cm \times
V(1_q^+,2^+,\ldots,(n-2)^+,(n-1)^-, (-K_{1 \ldots (n-1)})_{\qb}^{ -})
\nonumber \\
&& \hskip 1.2 cm \null 
+ J_{\mu}((K_{1 \ldots n})^- ; P_\V )
\frac{i}{K_{1 \ldots n}^2}
%\nonumber \\
%&& \hskip 2.5 cm \times \null 
V(1_q^+,2^+,\ldots,(n-2)^+,(n-1)^-,n_{\qb}^-, (-K_{1\ldots n})^{+}) \ ,
\label{NMHVCurrent}
\end{eqnarray}
} %% end footnotesize
where the momentum of the vector boson is $P_\V = - K_{1\ldots n}$.
The explicit values of the current vertices are obtained from
eqns.
(\ref{BasicFermionicCurrent}) and (\ref{BasicGluonicCurrent})
by relabeling the
arguments.  Other NMHV helicity configurations are only a bit more
complicated.
In ref.~\cite{MHVRecursive}, Bena and two of the authors introduced
a recursive reformulation of the CSW rules, useful when increasing
the number of negative helicity legs.  
Indeed, an analogous recurrence relation also applies to the
vector-boson currents considered here.

%%%%%%%%%%%%%%%%%%%%%%%%%%%%%%%%%%%%%%%%%%%%%%%%%%%%%%%

\section{Conclusions}
\label{ConclusionSection}

The twistor-inspired computational approach presented by Cachazo,
Svr\v{c}ek, and Witten \cite{CSW}, and very recently demonstarted by Risager
\cite{CSWdemo}, is among the novel ways of computing tree amplitudes in 
massless gauge theories, including of course QCD.  
In this talk, I have discussed the main issue of \cite{Bern:2004ba},
where, with Z. Bern, D. Forde, and D. Kosower,
addressing the question of computing
amplitudes containing both colored and non-colored particles,
we have shown how to incorporate
an additional vector leg coupling to an arbitrary source into the
CSW approach.  
The currents we have constructed can be used directly in the
computation of processes producing electroweak vector bosons.
The structure of the CSW construction implies that
that a similar approach can be used to build 
multi-$W$s currents.

In outlook, novel techniques dealing directly
with on-shell objects,  
like the CSW \cite{CSW} and the BCFW \cite{BCFW} approaches, 
relying on general properties of complex analysis, and 
exploiting the {\it recursive behaviour} of scattering
amplitudes, are establishing themselves as suitable tools
\cite{Cachazo:2005ga, Dixon:2005sp} for computing
massless and massive multi-legs tree-level \cite{TreeApplications}
and one-loop QCD (and beyond) amplitudes \cite{LoopApplications}.

%%%%%%%%%%%%%%%%%%%%%%%%%%%%%%%%%%%%%%%%%%%%%%%%
%% BACKMATTER
%%%%%%%%%%%%%%%%%%%%%%%%%%%%%%%%%%%%%%%%%%%%%%%%



\bibliographystyle{aipproc}   % if natbib is available

%\bibliography{sample}

%  \typeout{******************************************}
%  \typeout{** Please run "bibtex \jobname" to optain}
%  \typeout{** the bibliography and then re-run LaTeX}
%  \typeout{** twice to fix the references!}
%  \typeout{******************************************}

%\end{document}

%%%%%%%%%%%%%%%%%%%%%%%%%%%%%%%%%%%%%%%%%%%
%% The following lines show an example how to produce a bibliography
%% without the help of the BibTeX program. This could be used instead
%% of the above.
%%%%%%%%%%%%%%%%%%%%%%%%%%%%%%%%%%%%%%%%%%%

% \endinput
%%
%% End of file `template-6s.tex'.
\end{document}